\documentclass[aps,prl,twocolumn,showpacs,referee,superscriptaddress,showpacs, amsmath, amssymb]{revtex4}
\usepackage{graphicx}
\usepackage{dcolumn}
\usepackage{bm}
\begin{document}
\def\btt#1{{\tt$\backslash$#1}}
\def\BibTeX{\rm B{\sc ib}\TeX}
\draft

\title{Hydromagnetic Instability in plane Couette Flow} 
\author{Alfio Bonanno}
\affiliation{ INAF, Osservatorio Astrofisico di Catania,
           Via S.Sofia 78, 95123 Catania, Italy} 
\affiliation{INFN, Sezione di Catania, Via S.Sofia 72,
           95123 Catania, Italy} 
\author{Vadim Urpin}
\affiliation{ INAF, Osservatorio Astrofisico di Catania,
           Via S.Sofia 78, 95123 Catania, Italy} 
\affiliation{A.F.Ioffe Institute of Physics and Technology, 
           194021 St. Petersburg, Russia}

\date{\today}
\begin{abstract}
We study the stability of a compressible magnetic plane Couette flow and 
show that compressibility profoundly alters the stability properties if the
magnetic field has a component perpendicular to the direction of flow.
The necessary condition of a newly found instability can be satisfied in 
a wide variety of flows in laboratory and astrophysical conditions. The 
instability can operate even in a very strong magnetic field which entirely 
suppresses other MHD instabilities. The growth time of this instability 
can be rather short and reach $\sim 10$ shear timescales.  
\end{abstract}


\maketitle

\section{Introduction}

Instabilities of the magnetized shear flows  play an important role in 
enhancing transport processes in various astrophysical bodies and laboratory 
experiments. Shear flows are rather common in astrophysics, and the study of 
their stability properties is important for the understanding of many 
phenomena in stars, jets, accretion flows, galaxies, etc. 

Likely, the best studied shear flow is differential rotation. It is well 
known since the classical papers by Velikhov \cite{vel59} and Chandrasekhar 
\cite{chandra60} that a differentially rotating flow with a negative angular 
velocity gradient and a weak magnetic field can be subject to the 
magnetorotational instability. This instability has been analyzed in detail 
in several astrophysical contexts (\cite{fricke69,ach78,bh91}) because it can 
be responsible for transport of the angular momentum in various objects. In 
accretion disks, this instability is also well studied by numerical 
simulations in both linear and non-linear regimes (see, e.g., \cite{bra95,
haw95,mat95}). Astrophysical applications rise great interest in trying to 
study this instability in laboratory \cite{rued01,ji01,rrb04}. The 
experiments, however, are complicated because very large rotation rates 
should be achieved. 

The plane Couette flow is another example of shear flows well studied in
laboratory conditions. The pioneering work on the stability of a magnetized 
plane Couette flow has been done by Velikhov \cite{veli59} who obtained that 
a longitudinal magnetic field with the strength $\geq 0.1 V_0 \sqrt{4 \pi 
\rho}$ has to stabilize the flow; $V_0$ is the velocity in the center of the 
channel and $\rho$ is the fluid density. A sufficient  condition of the ideal
instability in a parallel magnetic field has been considered by Chen \&
Morrison \cite{chen91}. They argued that the magnetic field can provide
a destabilizing effect such as the flow, which is stable in the absence of a
magnetic field, can be driven unstable by a relatively weak magnetic field.
Also, they found that although strong magnetic shear can stabilize shear flow,
there exist a range of magnetic shear that causes destabilization. The linear 
stability properties of dissipative shear flow in a parallel magnetic field 
have been considered by Lerner \& Knobloch \cite{ler85}. The authors argued 
that misaligned linear perturbations can exhibit enhanced decay in such 
dissipative flows. Stability of incompressible flow in a transverse magnetic 
field has been studied by Takashima \cite{tak96, tak98} who found that
there exist both the stationary and traveling modes of instability.   

Note that many previous stability analyses have adopted the Boussinesq 
approximation, and have therefore neglected the effect of compressibility. 
This is allowed if the magnetic field strength is essentially subthermal, 
and the sound speed is much greater than the Alfv\'en velocity, $c_{s} \gg 
c_{A}$ but often this cannot be realized in real astrophysical conditions 
and in many numerical simulations. As it was shown by Bonanno \& Urpin
\cite{bon06}, the compressibility profoundly alters the stability properties
of shear flows. The number of new instabilities may occur in a compressible 
flow if the magnetic field has a component perpendicular to the flow.
Bonanno \& Urpin \cite{bon06} have considered the particular case of
differentially rotating flows but, likely, the shear-driven instabilities
are typical for other shear flows as well. In this paper, we show that the 
same sort of MHD instabilities can occur also in a plane Couette flow if the 
magnetic field has a transverse component. The instability can arise even in 
a sufficiently strong magnetic field that suppresses other MHD instabilities. 
Stability analysis done in this paper will hopefully prove to be a useful 
guide in understanding various numerical simulations that explore the 
nonlinear development of instabilities and their effects on the resulting 
turbulent state of shear flows.

\section{Basic equations}

Consider a plane Couette flow with the velocity $\vec{V} = V(z) 
\vec{e}_{y}$ where $x$, $y$, and $z$ are the Cartesian coordinates; 
$\vec{e}_{x}$, $\vec{e}_{y}$, and $\vec{e}_{z}$ are the unit vectors. 
For the sake of simplicity, we assume shear to be linear, $V(z) =
V_{0} + z V'$, where $V_0$ and $V'$ are constant.

{We restrict ourselves to an inviscid fluid}. The equations of 
compressible MHD read in this case  
\begin{eqnarray}
\dot{\vec{v}} + (\vec{v} \cdot \nabla) \vec{v} = - \frac{\nabla p}{\rho} 
+ \frac{1}{4 \pi \rho} (\nabla \times \vec{B}) \times \vec{B}
+ \frac{\vec{F}}{\rho}, 
\end{eqnarray}
\begin{equation}
\dot{\rho} + \nabla \cdot (\rho \vec{v}) = 0, 
\end{equation}
\begin{equation}
\dot{p} + \vec{v} \cdot \nabla p + \gamma p \nabla \cdot 
\vec{v} = 0,
\end{equation}
\begin{equation}
\dot{\vec{B}} - \nabla \times (\vec{v} \times \vec{B}) + \eta
\nabla \times (\nabla \times \vec{B}) = 0,
\end{equation}
\begin{equation}
\nabla \cdot \vec{B} = 0. 
\end{equation} 
Our notation is as follows: $\rho$ and $\vec{v}$ are the density and 
fluid velocity, respectively; $p$ is the gas pressure; $\vec{B}$ 
is the magnetic field, $\eta$ is the magnetic diffusivity, and $\gamma$ 
is the adiabatic index; $\vec{F}$ is a scalar force introduced in order to
provide hydrostatic equilibrium in the basic state. For the sake of 
simplicity, the flow is assumed to be isothermal. 

The basic state on which the stability analysis is performed is assumed 
to be quasi-stationary with the magnetic field that has non-vanishing 
components in all directions, $\vec{B}= (0, B_{y}(z), B_{z})$. Generally, 
a quasi-stationary basic state in such a magnetic shear flow can be achieved 
only if dissipative effects are taken into account. In the basic state, 
the magnetic field $\vec{B}$ should satisfy the stationary induction 
equation
\begin{equation}
- \eta \Delta \vec{B} = \vec{e}_{y} V' B_{z}. 
\end{equation}
Since $\vec{B}$ depends only on the $z$-coordinate, we have
\begin{equation}
\frac{d^2 B_{y}}{dz^2} = - \frac{V' B_{z}}{\eta}.
\end{equation}
Integrating this equation, we obtain
\begin{equation}
B_{y} = - \frac{V' B_{z} z^2}{2 \eta} + B_{0y}' z + B_{0y},
\end{equation}
where $B_{0y}'$ and $B_{0y}$ are constant. We can choose the boundary 
conditions in such a way that $B_{0y}'=0$ that corresponds to the absence 
of electric currents at the low boundary $z=0$. We will assume that the 
longitudinal magnetic field at the low boundary is much stronger than 
$V' B_{z} d^2/ 2 \eta$ where $d$ is the thickness of the Couette flow. 
Then, $B_{y} \approx B_{0y}$, and one can neglect the change of $B_{y}$ 
across the basic flow when considering the behaviour of small perturbations. 

{ We assume also that the basic state satisfies the condition of 
hydrostatic 
equilibrium in the $x$- and $z$-directions. In the $x$-direction, this 
condition is satisfied always. For the chosen magnetic field, hydrostatic
equilibrium in the $z$-direction yields 
\begin{equation}
\frac{d p}{d z} + \frac{1}{4 \pi} \frac{d}{d z} (B_y^2) + F_z = 0. 
\end{equation}
Eq.~(9) can be satisfied if $p=$const and the vertical component of the
Lorentz force is balanced by a scalar force in the basic state,
\begin{equation}
F_z = \frac{1}{4 \pi} \frac{d B_y^2}{d z}.
\end{equation}
Note that, generally, there is no hydrostatic equilibrium in the $y$-direction
in our model, but departures from equilibrium are small and can lead only to 
a very slow change of the basic state. For example, the Lorenz force changes
the basic velocity profile $V(z)$ in accordance with the $y$-component of the
momentum equation,
\begin{equation}
\dot{v}_y = \frac{B_z B_y'}{4 \pi \rho} = - \frac{V' B_z^2 z}{4 \pi \rho} =
- \frac{c_{Az}^2 V' z}{\eta},
\end{equation} 
where $c_{Az}^2 = B_z^2/ 4 \pi \rho$. Integrating this expression, we obtain
\begin{equation}
v_y (t) \approx V(z) - \frac{c_{Az}^2 V' z t}{\eta}.
\end{equation}
The Lorentz force changes essentially the initial velocity profile on the 
timescale 
\begin{equation}
\tau_{0y} \sim \frac{\eta V(z)}{c_{Az}^2 V' z} \sim \frac{\eta}{c_{Az}^2}
\end{equation}
(we assume $V(z) \sim zV'$). Therefore, one can neglect this departure from
hydrostatic equilibrium if the growth rate of instability, $\sigma$, is 
greater than $1/ \tau_{0y}$, or
\begin{equation}
\sigma \gg \frac{c_{Az}^2}{\eta}.
\end{equation}
Under this condition, the chosen basic state can be considered as 
quasi-stationary. We will show that this condition is satisfied in many 
cases of interest.}

We consider the stability of perturbations with the spacetime dependence 
$\propto f(z) \exp ( \sigma t )$. Small perturbations will be indicated by 
subscript 1, while unperturbed quantities will have no subscript. Then, 
the linearized MHD-equations read  
\begin{eqnarray}
\sigma \vec{v}_{1} + \vec{e}_{y} V' v_{1z}    
= \frac{\rho_{1}}{\rho^2} \left[ \nabla p - \frac{1}{4 \pi}
(\nabla \! \times \! \vec{B}) \! \times \! \vec{B} + \vec{F} \right]
\nonumber \\
- \frac{\nabla p_{1}}{\rho} \!
+ \frac{1}{4 \pi \rho} [ (\nabla \times \vec{B}_{1}) \times \vec{B} +
(\nabla \times \vec{B}) \times \vec{B}_{1} ],  
\end{eqnarray}
\begin{equation}
\sigma \rho_{1} + \nabla ( \rho \vec{v}_{1}) = 0, 
\end{equation}
\begin{equation}
\sigma p_{1} + \gamma p (\nabla \cdot \vec{v}_{1}) = 0, 
\end{equation}
\begin{equation}
\sigma \vec{B}_{1} = \vec{e}_{y} V' B_{1z }  
+ \nabla \times ( \vec{v}_{1} \times \vec{B}) + \eta \Delta 
\vec{B}_{1}, 
\end{equation}
\begin{equation}
\nabla \cdot \vec{B}_{1} = 0.
\end{equation}
This set of equations determines the behaviour of small perturbations.

\section{Criteria of instability}

A general set of Eqs.~(15)-(19) can be substantially simplified under our
assumptions regarding the basic state. As it was mentioned, small departures 
from hydrostatic equilibrium in the basic state can not influence the 
behavior of perturbations if inequality (14) is satisfied. Therefore, the 
term proportional to $\rho_1/\rho$ on the r.h.s. of Eq.~(15) can be neglected 
since it is proportional to small departures from hydrostatic equilibrium.
From Eq.~(19), we have $\partial B_{1z}/ \partial z = 0$ and, hence, $B_{1z}
=0$. Then, the $x$-, $y$-, and $z$-components of the momentum equation are
\begin{eqnarray}
\sigma v_{1x} = \frac{B_z}{4 \pi \rho} \frac{\partial  B_{1x}}{\partial z}, \\
\sigma v_{1y} + V' v_{1z} = \frac{B_z}{4 \pi \rho} 
\frac{\partial B_{1y}}{\partial z}, \\
\sigma v_{1z} = \frac{c_s^2}{\sigma} \frac{\partial^2 v_{1z}}{\partial^2 z}
- \frac{1}{4 \pi \rho} \frac{\partial}{\partial z} (B_{y} B_{1y} ).
\end{eqnarray}
Eq.~(23) yields for the $x$- and $y$-components of the magnetic field
\begin{eqnarray}
\left( \sigma - \eta \frac{\partial^2}{\partial z^2} \right) B_{1x} =
B_z \frac{\partial v_{1x}}{\partial z} , \\
\left( \sigma - \eta \frac{\partial^2}{\partial z^2} \right) B_{1y} =
B_z \frac{\partial v_{1y}}{\partial z} -  
\frac{\partial}{\partial z} (B_{y} v_{1z}).
\end{eqnarray}
Combining now Eqs.~(21), (22), and (24) and taking into account that $B_{y}
\gg d B_{y}'$, we can obtain the equation that contains only
perturbation $v_{1z}$,
\begin{eqnarray}
\left\{ \! \left[ \sigma \left( \! \sigma \! - \! \eta 
\frac{\partial^2}{\partial z^2} 
\! \right)
\! - \! c_{Az}^2 \frac{\partial^2}{\partial z^2} \! \right] \left( \! 
\sigma^2 -
c_s^2 \frac{\partial^2}{\partial z^2} + \frac{B_y}{B_z} \sigma V' \! \right) 
\right. \nonumber \\
\left. - \frac{B_y}{B_z} \sigma^2 \left[ V' \left( \sigma - 
\eta \frac{\partial^2}{\partial z^2} \right) + \frac{B_y}{B_z} c_{Az}^2
\frac{\partial^2}{\partial z^2} \right] \right\} v_{1z} =0,
\end{eqnarray}
where $c_s^2 =\gamma p/\rho$. This equation
can be solved easily since all coefficients are approximately constant in our
model. To solve Eq.~(25), one needs the boundary conditions. Note that the 
eigenvalues are not very sensitive to the boundary conditions. Therefore,
we choose the simplest model conditions and assume that $v_{1z}=0$ at $z=0$
and $z=d$. Then, $v_{1z} \propto \sin qz$ where $q = \pi n/d$ and $n$ is
integer. From Eq.~(25), we have the following dispersion equation for the
fundamental mode ($n=1$)
\begin{eqnarray}
\sigma^4 + \sigma^3 \omega_{\eta} + \sigma^2 (\omega_s^2 + \omega_m^2)+ \sigma
(\omega_{BV}^3 + \omega_s^2 \omega_{\eta})  \nonumber \\ 
+ \omega_{Az}^2 \omega_s^2 = 0,
\end{eqnarray}
where $\omega_{\eta}= \eta q^2$, $\omega_s = c_s q$, $\omega_{Az} = c_{Az} q$,
$\omega_{BV}^3 = q^2 c_{Az} c_{Ay} V'$, and $\omega_m^2 =q^2 (c_{Ay}^2 +
c_{Az}^2)$; $c_{Ay}^2 =B_y^2/ 4 \pi \rho$. This equation describes fast and 
slow magnetosonic waves modified by shear.

The conditions under which Eq.~(26) has unstable solutions can be obtained 
by making use of the Routh-Hurwitz theorem (see \cite{hen}, \cite{alek}). 
In the case of the dispersion equation of a fourth order, 
the Routh-Hurwitz criteria are written, for example, in \cite{mir}.
According to these criteria, Eq.~(26) has unstable solutions if one 
of the following inequalities is fulfilled
\begin{eqnarray}
\omega_{\eta} < 0 \;, \;\;\;  \omega^{2}_{Az} \omega^{2}_{s} < 0 \;,\;\; \\
\omega^{3}_{B V} - \omega_{\eta} \omega_m^2  > 0 \;, \\
\omega^{6}_{B V} + \omega^{3}_{B V} \omega_{\eta} (\omega_s^2 - \omega_m^2)
- \omega_{\eta} \omega_s^2  \omega_{Ay}^2 > 0 \;,
\end{eqnarray}
where $\omega_{Ay}^2 = c_{Ay}^2 q^2$. Two conditions (27) never apply because 
$\omega_{\eta}$, $\omega_{Az}^2$, and $\omega_s^2$ are positive. In the limit 
of small magnetic diffusivity, Eqs.~(28)-(29) are equivalent to
\begin{equation}
\omega_{B V}^3 \neq 0
\end{equation}
that is the generalization of the condition derived by Bonanno and Urpin 
\cite{bon06} for differentially rotating flows. Apart from shear, condition 
(30) requires non-vanishing $y$- and $z$-components of the magnetic field. 
The direction of $\vec{B}$ and the sign of $V'$ are insignificant, and the 
instability may occur for both positive and negative $V'$. Note that the 
instability given by Eq.~(30) can arise even in a very strong field.

Consider criteria (28)-(29) in the case when dissipation can not be neglected.
Condition (28) can be rewritten as
\begin{equation}
\frac{B_z}{B_y} > 2 \pi^2 \left( \frac{2 \eta}{d^2 V'} \right) \left(1 + 
\frac{B_z^2}{B_y^2} \right).
\end{equation}
As it was mentioned, our consideration is valid only if the condition
$B_y > V' B_z d^2/ 2 \eta$ is satisfied (see Eq.~(8)) that is equivalent
to
\begin{equation}
\frac{B_z}{B_y} < \frac{2 \eta}{d^2 V'}.
\end{equation} 
Since inequalities (31) and (32) are incompatible in the chosen longitudinal
field, criterion (28) can not be fulfilled and, hence, $\omega_{B V}^3 -
\omega_{\eta} \omega_m^2 <0$.  

Since $\omega_{B V}^3 - \omega_{\eta} \omega_m^2 <0$ in the considered flow, 
we can transform criterion (29) into
\begin{equation}
\omega_{B V}^3 + \omega_{\eta} \omega_s^2  + \frac{\omega_{\eta}^2 
\omega_s^2 \omega_{Az}^2}{\omega_{B V}^3 -\omega_{\eta} \omega_m^2} < 0.  
\end{equation}
Taking into account Eq.~(32), we can estimate $|\omega_{B V}^3 -
\omega_{\eta} \omega_m^2| \sim \omega_{\eta} q^2 c_{Ay}^2$. Then, the last
term on the l.h.s. is of the order of $\omega_{\eta} \omega_s^2 (B_z/B_y)^2$
and can be neglected compared to the second term. Hence, criterion (33) is
approximately equivalent to
\begin{equation}
\omega_{B V}^3 + \omega_{\eta} \omega_s^2 < 0.  
\end{equation}
This condition can be fulfilled only if 
\begin{equation}
B_z B_y V' < 0
\end{equation}  
that is the necessary condition of instability. In accordance with this 
condition, the imposed longitudinal field should have the same direction 
as the field stretched from $B_z$. If inequality (35) is satisfied, then
the instability arises if $|\omega_{BV}^3| > \omega_{\eta} \omega_s^2$, or
\begin{equation}
\frac{c_{Ay}^2}{c_s^2} > 2 \pi^2  \frac{B_y}{B_z} \frac{2 \eta}{d^2 V'}.
\end{equation}
This inequality can be fulfilled in a wide variety of strongly magnetized
flows where the magnetic pressure is greater than the thermal pressure.
Eqs.~(35) and (36) determine the necessary and sufficient conditions of
shear-driven instability in a Couette flow.

\section{The growth rate of instability}

Since the necessary condition of instability is given by Eq.~(35), we consider
the roots of Eq.~(26) only in the case of negative $\omega_{BV}^3$ when
$\omega_{BV}^3 = - |\omega_{BV}^3|$. To calculate the growth rate it is 
convenient to introduce dimensionless quantities
\begin{eqnarray}
\Gamma= \frac{\sigma}{|V'|} \;,\;\; \xi = 
\frac{4 \pi^2 \eta}{d^2 |V'|} \;, \;\; \epsilon=\frac{B_z}{B_y} \;,
\nonumber \\ 
\alpha = \frac{q^2 c_{Ay}^2}{|V'|^2} \;, \;\; \beta=\frac{c_s^2}{c_{Ay}^2}
\;. \nonumber
\end{eqnarray}
Then, Eq.~(26) becomes 
\begin{equation}
\Gamma^{4} + \Gamma^{3} \xi + \Gamma^{2} \alpha (1 + \epsilon^2 + \beta) + 
\Gamma \alpha  ( \beta \xi - \epsilon) + \epsilon^2 \beta \alpha^2 = 0.
\end{equation}
This equation was solved numerically for different values of the parameters
by computing the eigenvalues of the matrix whose characteristic 
polynomial is given by Eq.~(26) (see \cite{press}, for details). Moreover
it is not difficult to see that in order to satisfy the constrain Eq.(8)
we must choose $\epsilon \ll 1$.

\begin{figure}
\includegraphics[width=9cm]{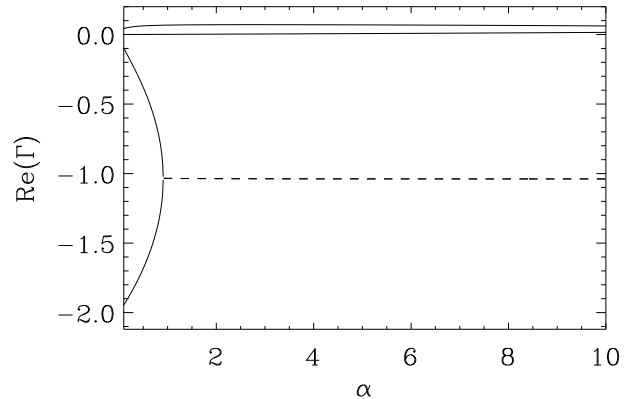}
\caption{The dependence of the real part of $\Gamma$ on 
$\alpha$ for $\beta =0.01$, $\epsilon = 0.1$, and $\xi=2$. Solids lines show 
the growth rate and frequency of the real roots, and the dashed line 
corresponds to the complex root.}
\end{figure}

In Fig.~1, we plot the dependence of real roots and real part of complex
roots on $\alpha$ for $\epsilon =0.1$, $\xi =2$ and $\beta=0.01$. The solid 
lines show roots when they are real, and the dashed line show the real part 
of complex roots. Our calculations clearly indicate that two real roots are 
positive for the considered parameters and, hence, there should exist a new 
shear-driven instability. The pair of complex roots split into a pair of 
real ones at $\alpha \approx 0.9$, but these roots always correspond to 
stable modes. One unstable root is rather large with the growth rate $\sim 
0.05-0.08 V'$, and another one is typically about 10 times smaller. For 
these roots, the growth rate varies very slowly with the parameter $\alpha$. 
Only if the Alf\'ven frequency is smaller than the characteristic shear 
frequency and $\alpha < 1$, the growth rate of the most unstable mode 
decreases. Note that the considered instability occurs at a very large 
magnetic pressure that exceeds the gas pressure by two orders of magnitude.   

\begin{figure}
\includegraphics[width=9cm]{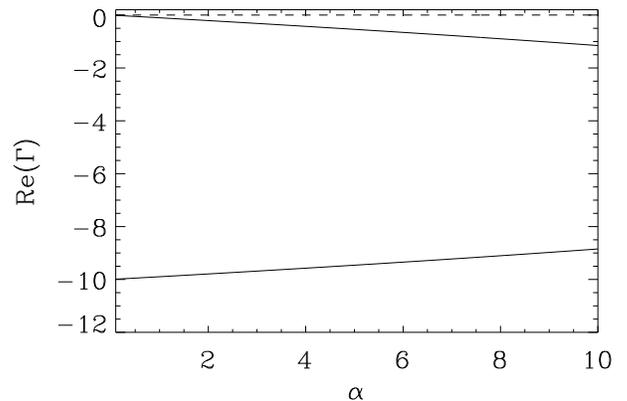}
\caption{The same as in Fig.~1 but for $\xi = 10$.}
\end{figure}

In Fig.~2, we plot the same dependence as in Fig.~1 but for $\xi=10$. The
higher value of $\xi$ corresponds to a larger magnetic viscosity and, hence,
to a stronger dissipation of perturbations. Due to this, the instability
turns out to be suppressed. Indeed, all roots are either negative or have
a negative real part. In this case, complex conjugate roots have a very
small negative  part, but real roots dissipate much more rapidly. Note that
the critical value $\xi$ that discriminate between stable and unstable
flows is $\sim 10$, and the instability occurs if $\xi < 10$. For example,
the growth rate can reach $\sim 0.04 V'$ in a flow with $\xi=5$.

\begin{figure}
\includegraphics[width=9cm]{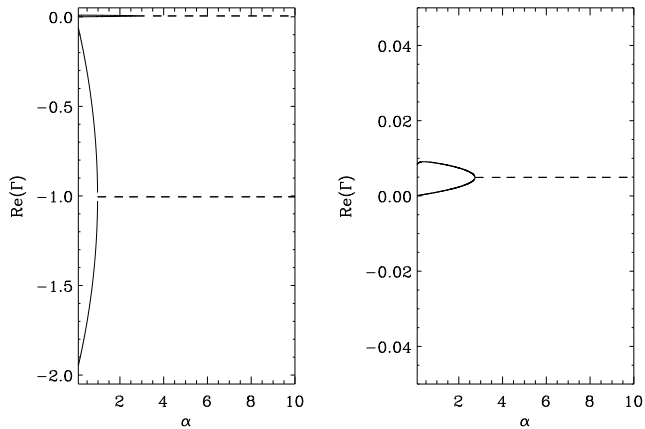}
\caption{The same as in Fig.~1 but for $\epsilon = 0.03$.}
\end{figure}

Fig.~3 shows the dependence of a real part of $\Gamma$ on $\alpha$ for 
$\beta=0.01$, $\varepsilon=0.03$, and $\xi=2$. The right panel shows the 
behavior of roots at the top left region of the left panel where roots are 
small. Comparing with Fig.~1, it is seen that a decrease of the ratio 
$\epsilon = B_z/B_y$ results naturally in a smaller growth rate. This 
dependence is qualitatively clear since the considered instability is due to 
the presence of a transverse field component in a flow and, therefore, a 
decrease of this component leads to a weaker instability. In the considered 
range of $\alpha$, both oscillatory and non-oscillatory modes can arise. Two 
non-oscillatory modes are unstable if $\alpha < 2.8$. After merging, this
couple forms a pair of complex conjugate modes that are unstable if 
$\alpha > 2.8$. The growth rate of one non-oscillatory is larger than that 
of oscillatory modes and can reach $\sim 0.01 V'$. The growth rate of 
oscillatory modes is a factor $\sim 2$ smaller. Note that another couple
of modes is always stable.  

\begin{figure}
\includegraphics[width=9cm]{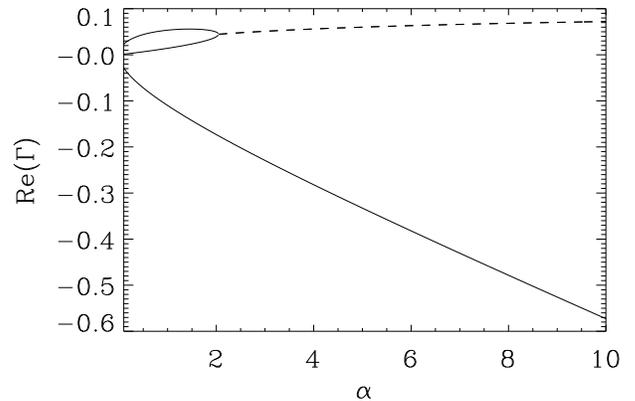}
\caption{The dependence of a real part of $\Gamma$ on 
$\alpha$ for $\epsilon =0.5$, $\xi = 30$, and $\beta=0.01$.}
\end{figure}

In Fig.~4, we plot the dependence of a real part of $\Gamma$ on $\alpha$
for $\epsilon=0.5$, $\xi=30$, and $\beta=0.01$. We show only three roots in
this figure since the fourth root has a large negative value, $\Gamma \sim
\xi \sim -30$. The increase of $\epsilon$ leads to a corresponding increase 
in the growth rate. Like the previous case, the instability can arise
either in oscillatory or non-oscillatory regimes. Two non-oscillatory modes 
are unstable if $\alpha < 2$. After merging at $\alpha \approx 2$, these 
real roots form a couple of complex conjugate roots that are unstable at 
$\alpha > 2$. Another pair of modes is always stable. The growth rate of
unstable modes is $\sim 0.07-0.1 V'$ and increases slightly with $\alpha$.
Note that oscillatory modes can grow faster than non-oscillatory ones in 
this case.

\begin{figure}
\includegraphics[width=9.0cm]{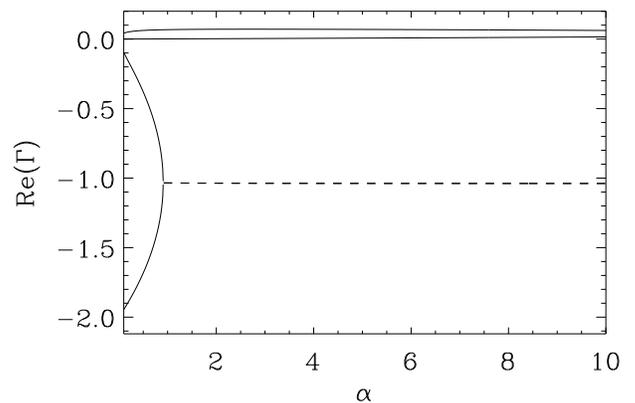}
\caption{The dependence of a real part of $\Gamma$ on $\alpha$ for 
$\epsilon=0.1$, $\xi = 2$, and $\beta=0.001$.}
\end{figure}

Fig.~5 shows the growth rate of instability for a very small value of 
$\beta=0.001$. This $\beta$ corresponds to the magnetic pressure 
approximately three orders of magnitude greater than the gas pressure.
Despite a very high magnetic pressure, the instability can still occur.
This is in an agreement with our analytic result that the instability 
should not be suppressed by a strong magnetic field.
The dependences in Fig.~5 are qualitatively very similar to those shown 
in Fig.~4. Two non-oscillatory modes are unstable in this case as well.
One unstable mode has a very small growth rate $\sim 0.01 V'$, but another 
one grows much faster, $\sigma \sim 0.1 V'$. The growth rate of the fastest
growing mode is even higher than in the case $\beta=0.01$ despite a strong
magnetic field.   


\section{Discussion}

To summarize then, we have considered the instability caused by shear in 
a compressible magnetized gas. To illustrate the main qualitative features 
of the instability associated to compressibility and shear, we analyzed a 
particular case of perturbations that depend on the vertical coordinate 
alone. The plane Couette flow with a non-vanishing transverse magnetic 
field turns out to be unstable even in this simplest case. The necessary 
condition of instability is $B_{z} B_{y} V' < 0$, and it can be easily 
satisfied in laboratory flows. Since the shear flow in the presence of a 
transverse magnetic field always stretches the longitudinal field satisfying 
the necessary condition (35), one can expect that the instability likely 
operates if $B_y$ is entirely generated by shear and $C_1=C_2=0$ in Eq.~(8). 
We consider this case elsewhere.  

{ The newly found instability is relatively slow: its growth rate 
reaches $\sim 0.1 V'$ and is small compared to the shear timescale, $1/V'$.
However, even this growth rate can be sufficient to generate hydrodynamic
motions in many real flows, for example, in astrophysics.} Basically, the 
growth rate is larger for non-oscillatory modes which are unstable at 
relatively not very large $\alpha$. The growth rate depends on the ratio of 
the magnetic and gas pressure, being smaller for a low ratio. 

The considered instability is related basically to shear and 
compressible properties of a magnetized gas. In the incompressible limit 
that corresponds to $c_{s} \rightarrow \infty$, we have from Eq.~(26)
\begin{equation}
\sigma^{2} + \sigma \omega_{\eta} + \omega_{Az}^{2} =0,
\end{equation}
{and the instability does not occur for the chosen perturbations. It can be not
the case, however, for perturbations of a more general form which depend
also on the $x$- or $y$-coordinates.} 

This new  instability can be either oscillatory or non-oscillatory, 
depending on the value of the ratio $2 \pi c_{Ay}/d V'$. Typically, the 
considered instability is non-oscillatory if $\alpha$ is not large and 
oscillatory in the opposite case. The critical $\alpha$ that determines
the transition between oscillatory and non-oscillatory regimes depends 
strongly on the parameters $\epsilon$, $\xi$, and $\beta$ and can vary
within a wide range.

One more important feature of the instability is associated with the 
dependence on the magnetic field strength. Generally, a sufficiently strong 
magnetic field can suppress instabilities of a shear flow. On the contrary, 
the instability discovered in our study cannot be suppressed even in very 
strong magnetic fields as it is seen from the criterion (40). All this 
comparison allows us to claim that our analysis demonstrates the presence 
of the new instability in compressible shear flows.

\vspace{0.5cm}
\noindent
{\it Acknowledgments.}
This research project has been supported by a Marie Curie Transfer of
Knowledge Fellowship of the European Community's Sixth Framework
Program under contract number MTKD-CT-002995.
VU thanks also INAF-Ossevatorio Astrofisico di Catania for hospitality.

\end{document}